\documentclass[aps,prl,showpacs,floats,twocolumn,floats,superscriptaddress,floatfix,longbibliography]{revtex4-1}
\usepackage{bm}
\usepackage{hyperref} 
\usepackage[usenames,dvipsnames]{xcolor}
\usepackage[normalem]{ulem}
\usepackage[utf8]{inputenc}
\usepackage{amsmath}
\usepackage{amsthm}
\usepackage{mathtools}
\usepackage{amsfonts}
\usepackage{amssymb}
\usepackage{appendix}
\usepackage{comment}

\usepackage{blindtext}
\usepackage[utf8]{inputenc}
\usepackage[T1]{fontenc}
\usepackage{graphicx}
\usepackage{float}
\usepackage{wrapfig}
\usepackage{bm}
\usepackage{braket}
\usepackage{grffile}
\usepackage[normalem]{ulem}
\usepackage[usenames,dvipsnames]{xcolor}
\usepackage{soul}

\newcommand{\be}{\begin{equation}}
\newcommand{\ee}{\end{equation}}

\newcommand{\bv}{\mathbf{v}}
\newcommand{\bV}{\mathbf{V}}
\newcommand{\bx}{\mathbf{x}}

\newcommand{\bR}{\mathbf{R}}
\newcommand{\bxstar}{\mathbf{x_{\star}}}

\begin{document}
%\linenumbers
\title{Hidden symmetry in passive scalar advected by 2D Navier-Stokes turbulence 
}
%%%%%%%%%%%%%%

\author{Chiara Calascibetta} \email{chiara.calascibetta@inria.fr}
\affiliation{Université Côte d'Azur, Inria, Calisto team, 06902 Sophia Antipolis, France}
%\affiliation{Department of Physics, University of Rome ``Tor
%Vergata'', Via della Ricerca Scientifica 1, 00133 Rome, Italy.} 
%\affiliation{INFN, Sezione di Roma ``Tor Vergata'', Rome, Italy.}
\author{Luca Biferale}
\affiliation{Department of Physics, University of Rome ``Tor
Vergata'', Via della Ricerca Scientifica 1, 00133 Rome, Italy.}
\affiliation{INFN, Sezione di Roma ``Tor Vergata'', Rome, Italy.}
\author{Fabio Bonaccorso}
\affiliation{Department of Physics, University of Rome ``Tor
Vergata'', Via della Ricerca Scientifica 1, 00133 Rome, Italy.}
\affiliation{INFN, Sezione di Roma ``Tor Vergata'', Rome, Italy.}
\author{Massimo Cencini}
\affiliation{Istituto dei Sistemi Complessi, CNR, Via dei Taurini 19, Rome, 00185, Italy.}
\affiliation{INFN, Sezione di Roma ``Tor Vergata'', Rome, Italy.}
\author{Alexei A. Mailybaev}
\affiliation{Instituto de Matemática Pura e Aplicada– IMPA, Rio de Janeiro, Brazil.}

%\date{\today}
\begin{abstract}
Here we show that passive scalars possess a \textit{hidden scaling
  symmetry} when considering suitably rescaled fields.  Such a symmetry
implies (i) universal probability distribution for scalar
multipliers and (ii) Perron-Frobenius scenario for the anomalous
scaling of structure functions. We verify these predictions with high
resolution simulations of a passive scalar advected by a 2D turbulent
flow in inverse cascade.
\end {abstract}

\maketitle

\begin{figure*}[t]
\centering 
\includegraphics[width=0.95\textwidth]{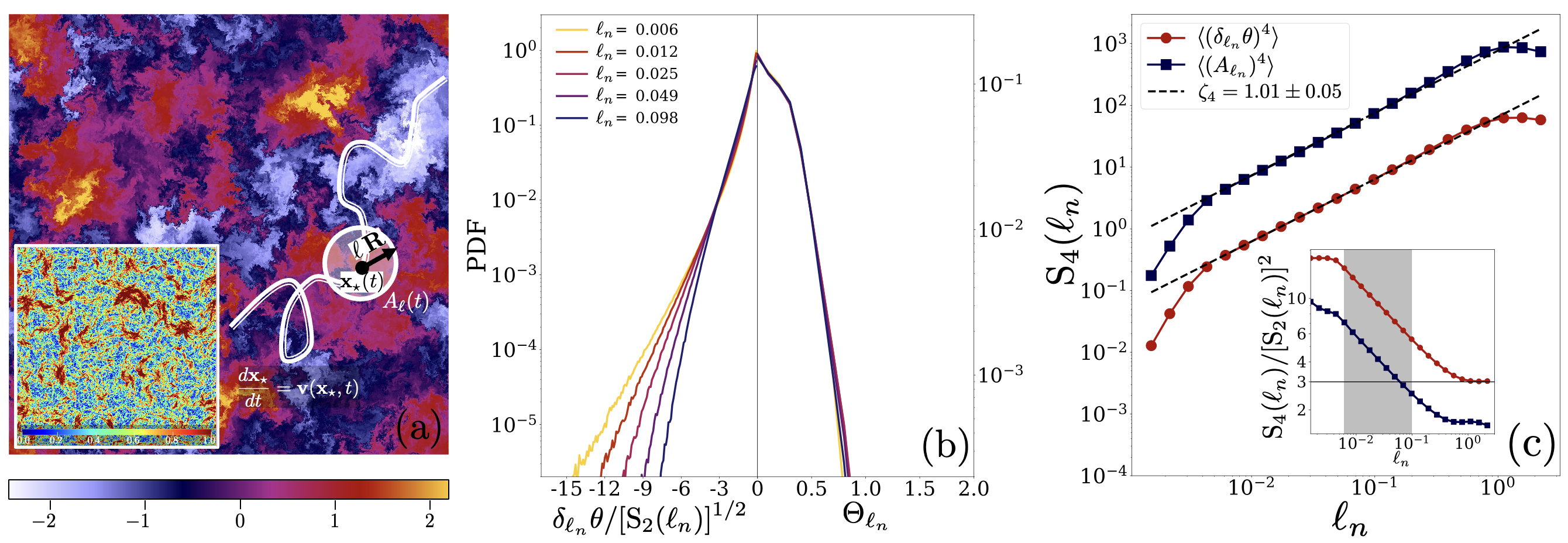}
\caption{Passive scalar in 2D inverse cascade: (a) Snapshot of the
  scalar field $\theta$ on a $4096^2$ grid and (inset) of the velocity
  modulus $|\bv|$. A sketch of the Quasi-Lagrangian frame is
  represented by a Lagrangian trajectory (white curve), which solves
  $\frac{d\bxstar}{dt} = \bv(\bxstar,t)$. At each time, scalar
  fluctuations $A_\ell(t)$ in Eq.~\eqref{eq:A_amplitude} are defined
  around the center $\bxstar(t)$. (b) Probability density functions
  (PDFs) of (negative tails) normalized scalar increments,
  $\delta_{\ell_n}\theta / [{\rm{S}}_{2}({\ell_n})]^{1/2}$ (with
  ${\rm{S}}_{p}({\ell_n}) = \langle (\delta_{\ell_n}\theta)^p
  \rangle$) and of (positive tail) the rescaled field,
  $\Theta_{\ell_n}$, at scale $\ell_n = 2^{-n}$, within the inertial
  range; notice that both fields are symmetric under a change of sign.
  The x and y-axes are adjusted accordingly in each case.  (c)
  Fourth-order structure function computed using scalar increments
  (red circles), ${\rm{S}}_{4}({\ell_n}) = \langle
  (\delta_{\ell_n}\theta)^4\rangle$, and amplitudes
  \eqref{eq:A_amplitude} (blue squares), ${\rm{S}}^A_{4}({\ell_n}) =
  \langle A_{\ell_n}^4 \rangle$. Dashed lines indicate fitted slope,
  $\zeta_4$, within the inertial range. The inset shows the kurtosis,
  ${\rm{S}}_{4}({\ell_n}) / [{\rm{S}}_{2}({\ell_n})]^2$, as a function
  of $\ell_n$ for both quantities. The shaded region highlights the
  inertial range, where the statistical analysis of this study
  focuses.}
\label{fig:field}
\end{figure*}

%\textit{Introduction.} 

Scalar fields transported by turbulent flows are encountered in many
natural and engineering settings, from atmospheric dynamics
\cite{pasquill1983atmospheric} to combustion
\cite{williams2018combustion}. Passive scalar statistics exhibit
intermittency with universal anomalous scaling exponents, independent
of the injection mechanism
\cite{shraiman2000scalar,falkovich2001particles}.  Kraichnan's
insigths that passive scalar in Gaussian and time-uncorrelated flows
can be intermittent~\cite{kraichnan1994anomalous} led to
breakthroughs~\cite{gawedzki1995anomalous,chertkov1996anomalous,shraiman1996symmetry}
that, exploiting the linearity of the advection equation, linked
intermittency to the anomalous scaling of zero modes, homogeneous
solutions of the scalar multipoint correlation functions.  Zero modes,
originally derived by perturbative expansions in the Kraichnan model
(see also \cite{pumir1997perturbation,antonov1999anomalous,kupiainen2007scaling}), are
statistically preserved structures of the Lagrangian
dynamics~\cite{falkovich2001particles}. Universality and statistical
preservation under Lagrangian dynamics were then verified in realistic
flows
\cite{celani2001statistical,celani2000universality,celani2001fronts},
such as scalars advected by 2D turbulence in the inverse cascade,
which is non intermittent \cite{boffetta2000inverse}. After more than
twenty years, however, the zero-mode picture of anomalous scaling did
not allow to advance our understanding of anomalous scaling in
Navier-Stokes (NS) 3D turbulence \cite{frisch1995turbulence} nor in
active scalars \cite{celani2004active}.

Recently, it was proposed that NS equations and, similarly, shell
models for turbulence possess a \textit{hidden symmetry} (HS)
\cite{mailybaev2022hidden,mailybaev2022shell,mailybaev2022hidden,mailybaev2022hidden2,mailybaev2023hidden}
(the existence of HS in turbulence was somehow conjectured in
\cite{she1994universal}), namely become scale invariant when
considering suitably (nonlinearly) rescaled fields, which are thus non
intermittent. Notably, HS provides theoretical support to the
universality of multiplier statistics, as conjectured by Kolmogorov
third hypothesis~\cite{kolmogorov1962refinement} (see also
\cite{chen2003kolmogorov}) and a way to link anomalous scaling to the
eigevalues of a Perron-Frobenious operator \cite{mailybaev2022hidden},
offering also support to the multiplicative view of intermittency and
the multifractal model \cite{frisch1985fully}. Very recently, HS was
extended to a shell-model version of Kraichnan’s
model~\cite{thalabard2024zero}, previously rationalized within the
zero modes framework \cite{wirth1996anomalous,benzi1997analytic}. Such
results suggest the possibility of bridging HS with zero modes.
%for explaining intermittency 
% passive scalars, at least within the framework of simplified models.

In this Letter, we demonstrate that realistic
models of passive scalar turbulence also possess the hidden symmetry. We do
so by considering a homogeneous and isotropic passive scalar advected
by 2D NS turbulence in the inverse cascade
\cite{celani2001statistical,celani2000universality,celani2001fronts}, for which
we numerically verify the universality of scalar multipliers,
extending the Kolmogorov third hypothesis to scalar turbulence, and the Perron-Frobenius approach to anomalous exponents.

%\textit{Passive scalar in 2D turbulence.}
We consider the advection-diffusion equation for a passive scalar field, $\theta$, in two dimensions
\begin{equation}\label{eq:theta}
\partial_t \theta + \bv \cdot \nabla \theta = \kappa \Delta \theta + f_\theta,  
\end{equation}
where $\kappa$ is the scalar diffusivity, and $f_\theta$ is a Gaussian forcing applied at large scales~\cite{celani2000universality}. The velocity field, $\bv$, obeys the incompressible 2D NS equations  
\begin{equation}\label{eq:NS}
\partial_t \bv + (\bv \cdot \nabla) \bv = -\nabla p + \nu \Delta  \bv - \beta \bv + \mathbf{f},  
\end{equation}
with $p$ the pressure, $\nu$ the kinematic viscosity, $\beta$
 a large-scale friction term preventing large-scale energy accumulation, due to an inverse energy cascade, which is
sustained by a small scale Gaussian forcing, $\mathbf{f}$. Appendix~A in End Matter details the numerical implementation. 

As well established in experiments and numerical
simulations~\cite{paret1998intermittency,boffetta2000inverse,boffetta2012two},
in 2D inverse cascade, the velocity field is scale invariant with
Kolmogorov scaling. In other words, within the inertial range, the
velocity field exhibits the space-time scaling symmetry $\bx,\,t,\,\bv
\mapsto \gamma \bx, \,\gamma^{2/3}t,\,\gamma^{1/3}\bv$ for $\gamma>0$,
which is admitted by Eq.~\eqref{eq:NS} with $\nu\!=\!\beta\!=\!\bm
f\!=\!0$. Actually, see below, such scale invariance applies
to velocity differences rather than the velocity field itself due to
the sweeping effect.  Even with scale invariant velocity fields, the
advected scalar exhibits intermittency
\cite{shraiman2000scalar,falkovich2001particles}. As a result, for passive
scalars, an extra relation $\theta \mapsto \gamma^\alpha \theta$, with
arbitrary $\alpha \in \mathbb{R}$, adds to the space-time scaling
symmetry, leading to a broader family of symmetries. Indeed, the
stationary probability measure exhibits statistical intermittency in
the form of multifractal scaling \cite{frisch1985fully}. Intermittency reveals
in the statistics of scalar increments \be
\label{eq:scalarincrements}
\delta_{\ell}\theta(\bx,\bR,t) = \theta(\bx + \ell \bR,t) -
\theta(\bx,t)\,, \ee usually used with a direction vector $\mathbf{R} = \mathbf{e}$. As found in
\cite{celani2000universality} and further illustrated in
Fig.\ref{fig:field}(b), the normalized probability density functions
(PDFs) of $\delta_{\ell}\theta$ fail to collapse at
different scales $\ell_n = 2^{-n}$ and deviate from Gaussianity.
The lack of self-similarity is
linked to the sharp gradients (cliffs) and smooth regions
(ramps) characterizing turbulent scalar transport \cite{shraiman2000scalar,celani2000universality}, see Fig.\ref{fig:field}(a).  
%The observed scaling exponents agree with previous studies 
% \cite{celani2000universality,celani2001statistical,celani2001fronts}, 
% where universality with respect to forcing mechanisms was demonstrated.
 
%\textit{Hidden symmetry for passive scalar.}  
We now introduce the
hidden symmetry framework, a general approach designed to transform an
intermittent field into a non intermittent one by identifying a
suitable change of variables
\cite{mailybaev2022hidden,mailybaev2022shell,mailybaev2022hidden2,mailybaev2023hidden,thalabard2024zero}. While
this framework can be applied to any intermittent field, here
we focus on the passive scalar in 2D turbulence and seek a transformation that
eliminates the aforementioned dependence on $\alpha$ in the scaling
symmetry, underlying the observed intermittency.
We first demonstrate that the Kolmogorov scaling symmetry of velocity
increments, $\delta_{\ell}\bv(\bx,\bR,t) \!=\! \bv(\bx+\ell \bR,t)\! -\!
\bv(\bx,t)$, can be reformulated as a  symmetry of the
equations of motion in the Quasi-Lagrangian (QL) frame, removing the sweeping effect. The approach is then exported
to the passive scalar field.  

Let the Lagrangian tracer $\bxstar(t)$,
solving $d\bxstar/dt = \bv(\bxstar,t)$, be the moving origin in the QL
formulation and fix a reference scale $\ell$  within the
inertial range (see sketch in Fig.\ref{fig:field}a). Together, these quantities define the rescaled QL
coordinates~\cite{belinicher1987scale}: \be\label{eq:R_tau} \bR =
({\bx - \bxstar(t)})/{\ell}, \quad \tau = {t}/{t_{\ell}}\,, \ee where
the characteristic time $t_\ell$ is given by \be\label{eq:tln_vln}
t_{\ell} = {\ell}/{v_{\ell}},\,\,\, v_{\ell} = v_{\ell_f}
\,\left({{\ell}}/{\ell_f}\right)^{1/3} \,,  \ee with $v_{\ell_f}$ the
characteristic velocity at the forcing scale $\ell_f$.  We then define
the rescaled QL velocity field \be\label{eq:QL_V} \bV_{\ell}(\bR,\tau)
= {\delta_{\ell}\bv(\bxstar(t),\bR, t)}/{v_{\ell}}\,.  \ee
Differentiating Eq.~\eqref{eq:QL_V} with respect to $\tau$ and applying
the incompressible 2D Euler equations ($\nu,\beta,\mathbf f = 0$ in Eq.\eqref{eq:NS}) for the inertial interval, yields:
\be\label{eq:V_eq} \partial_\tau \bV \!+\! \bV\cdot\nabla_\bR \bV\!
=\! -\nabla_\bR P + (\nabla_\bR P)_{\bR=0},\,\, \nabla_{\bR}\cdot \bV
\!=\! 0\,, \ee where $\nabla_{\bR}$ is the gradient in the space
$\bR$. The new pressure field $P(\bR,\tau)$ is determined by the
incompressibility condition, just as in the Euler system.
Equation~\eqref{eq:V_eq} is invariant under the space-time scaling
transformation \be\label{eq:QL_V_sym} \bR,\,\tau,\,\bV \mapsto
\gamma\bR,\,\gamma^{2/3}\tau,\,\gamma^{1/3}\bV\,, \ee which
corresponds to the change of scale $\ell \mapsto \ell / \gamma$, that
is why the subscript $\ell$ in $\bV$ was omitted.  The symmetry
(\ref{eq:QL_V_sym}) is the precise mathematical
formulation of the K41 theory, because the field $\bV(\bR,\tau)$
describes velocity increments and is unaffected by sweeping.

We now extend this approach to the scalar field.  The QL scalar
difference, $\delta_{\ell}\theta(\bxstar(t),\bR,t)$, has the same
intermittent statistics of the Eulerian increments. Thus to obtain a
non intermittent field we need a suitable (nonlinear) normalization
accounting for the local scalar activity.  To this aim we introduce
the scalar amplitude, \be\label{eq:A_amplitude}
\begin{gathered}
A_{\ell}(t) \coloneqq \Big(\frac{1}{\pi}\!\int_{|\bR| \le 1}\!\!\!\!\!\!\!\!\!{\big[\delta_{\ell}\theta(\bxstar(t),\bR,t)\big]^2 d^2\bR} \Big)^{1/2} \,,
\end{gathered}
\ee 
which we can  write as $\langle
\big[\delta_{\ell}\theta(\bxstar(t),\bR,t)\big]^2
\rangle^{1/2}_{\mathcal D_1}$, i.e. the root-mean-square scalar increment within the disc ${\mathcal
  D_1}\! =\!\{\bR: |\bR| \leq 1 \}$ centered at $\bxstar(t)$.  $A_{\ell}(t)$ quantifies the average scalar fluctuations at scale $\ell$ and  its moments ${\rm{S}}^A_{p}({\ell_n}) =\langle A_{\ell_n}^p \rangle$ scale as the standard structure functions, ${\rm{S}}_{p}({\ell_n}) =\langle (\delta_{\ell_n}\theta)^p
\rangle$, as shown in Fig.~\ref{fig:field}(c) for $p=4$. The inset displays the flatness for both cases with, marked in grey, the inertial range considered in all the analysis discussed in this Letter.
We now introduce the rescaled passive scalar field
\be
\label{eq:Theta_def} \Theta_{\ell}(\bR,\tau) \coloneqq
\frac{\delta_{\ell}\theta(\bxstar(t),\bR,t)}{A_{\ell}(t)}\,  \ee
that, as shown in Sec.~I of Supplementary Material (SM) \footnote{See {S}upplemental {M}aterial [url], for a step-by-step derivation of Eq.~(\ref{eq:PDE_Theta}), demonstration that the change of variable \eqref{eq:HS} is a symmetry for Eq.~\eqref{eq:PDE_Theta} and corresponds to a change of scale, derivation of Eq.~\eqref{eq:new_mult}, a detailed description of the approximations of the Perron-Frobenious (PF) operator and its implementation to obtain the anomalous exponents \eqref{eq:SF_scaling_PF} and on the extension of PF scenario to other structure functions}, obeys the equation, 
\be\label{eq:PDE_Theta} \partial_\tau
\Theta + \bV\cdot \nabla_{\bR}\Theta - \Theta \langle \Theta \bV \cdot
\nabla_{\bR}\Theta\rangle_{\mathcal{D}_1} = 0\,.  \ee Equation~\eqref{eq:PDE_Theta}
exhibits a hidden scaling symmetry under the change of variable (see Sec.~II of SM \cite{Note1}) \be
\label{eq:HS} \Theta(\bR,\tau) \mapsto
\frac{\Theta(\bR/\gamma,\tau/\gamma^{2/3})}{B_\gamma(\tau)}\, , \ee
which combines K41 space-time scaling (\ref{eq:QL_V_sym}) with the
nonlinear normalization 
$B_\gamma(\tau) =\langle \Theta^2(\bR/\gamma,\tau/\gamma^{2/3})\rangle^{1/2}_{\mathcal{D}_1}$.
%\be\label{eq:HSnorm} B_\gamma(\tau) =
%\Big(\frac{1}{\pi}\int_{|\bR| \le
%  1}{\Theta^2(\bR/\gamma,\tau/\gamma^{2/3}) d^2\bR} \Big)^{1/2}.  \ee
This change of variables corresponds to the transformation $\ell \mapsto \ell/\gamma$
(Sec.~III of SM \cite{Note1}), justifying the term
\textit{scaling} symmetry and the omission of subscript $\ell$ in
$\Theta$ and $\bV$ in Eq.\eqref{eq:PDE_Theta}.  The symmetry is
\textit{hidden}  because it only manifests in the rescaled
system: it emerges after normalizing the QL scalar increments by its local scalar amplitude, which defines the rescaled
field in Eq.\eqref{eq:Theta_def}.  Unlike the scaling
relation $\theta \mapsto \gamma^\alpha \theta$ for the original scalar
field, where  $\alpha$ is arbitrary, HS~(\ref{eq:HS}) has no free parameter.

Summarizing, the recovery of hidden scaling symmetry implies that the
field $\Theta$ is statistically self-similar: \be\label{eq:law}
\Theta_{\ell} \overset{\text{law}}{=} \Theta_{\ell'},\,\,\,
\textrm{for } \ell \textrm{ and } \ell' \textrm{ in the inertial
  interval}\,. \ee This is verified on the right of
Fig.~\ref{fig:field}(b), showing the collapse of the PDFs of
$\Theta_{\ell}$ for $\ell$ within the inertial range (compare with the
left of the figure).  Actually, Eq.\eqref{eq:law} has more general
consequences, as it implies that the statistics of any observable of
$\Theta$ is independent of the scale. As shown below, this
has important implications with respect to the validity of (i) the
third Kolmogorov hypothesis on the universality of multiplier
statistics \cite{kolmogorov1962refinement,chen2003kolmogorov} (whose
definition is here extended to the passive scalar field) and (ii) the
Perron-Frobenius framework for the anomalous scaling of structure
functions \cite{mailybaev2022hidden,thalabard2024zero}.

\begin{figure}[b!]
\centering
\includegraphics[width=0.47\textwidth]{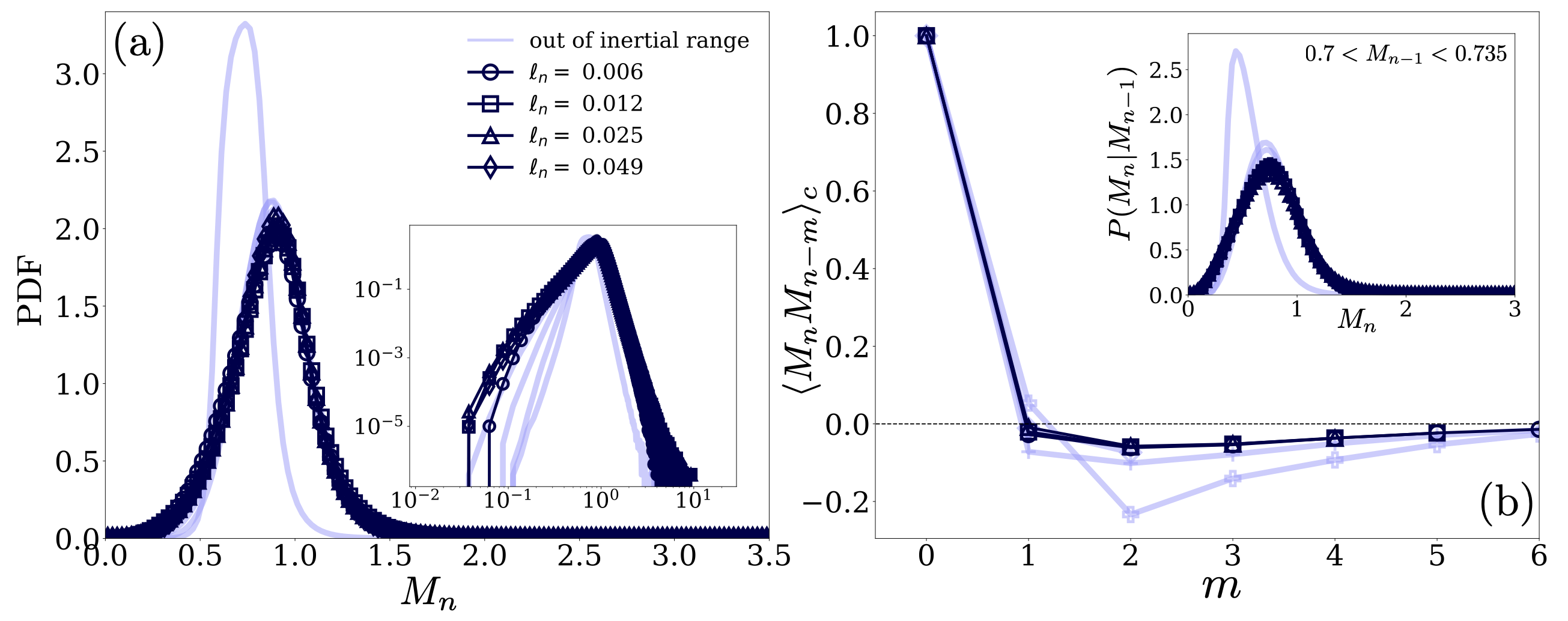}
\caption{(a) PDF of multipliers $M_n$ at different scales $\ell_n = 2^{-n}$; (inset) same data in log-log scale. (b) Correlation function of multipliers, $\langle M_n M_{n-m}\rangle_c = (\langle M_n M_{n-m}\rangle - \langle M_n \rangle^2 )/(\langle M_n^2 \rangle - \langle M_n \rangle^2)$. The inset presents the conditional distribution of multipliers between two consecutive scale, $\ell_n$ and $\ell_{n-1}$, showing $P(M_n | M_{n-1})$ for a fixed interval $0.7<M_{n-1}<0.735$. In all panels, dark/ligth blue curves indicate scales within/outside the inertial range.}
\label{fig:multpdf}
\end{figure}
\begin{figure*}[t!]
\centering
\includegraphics[width=1\textwidth]{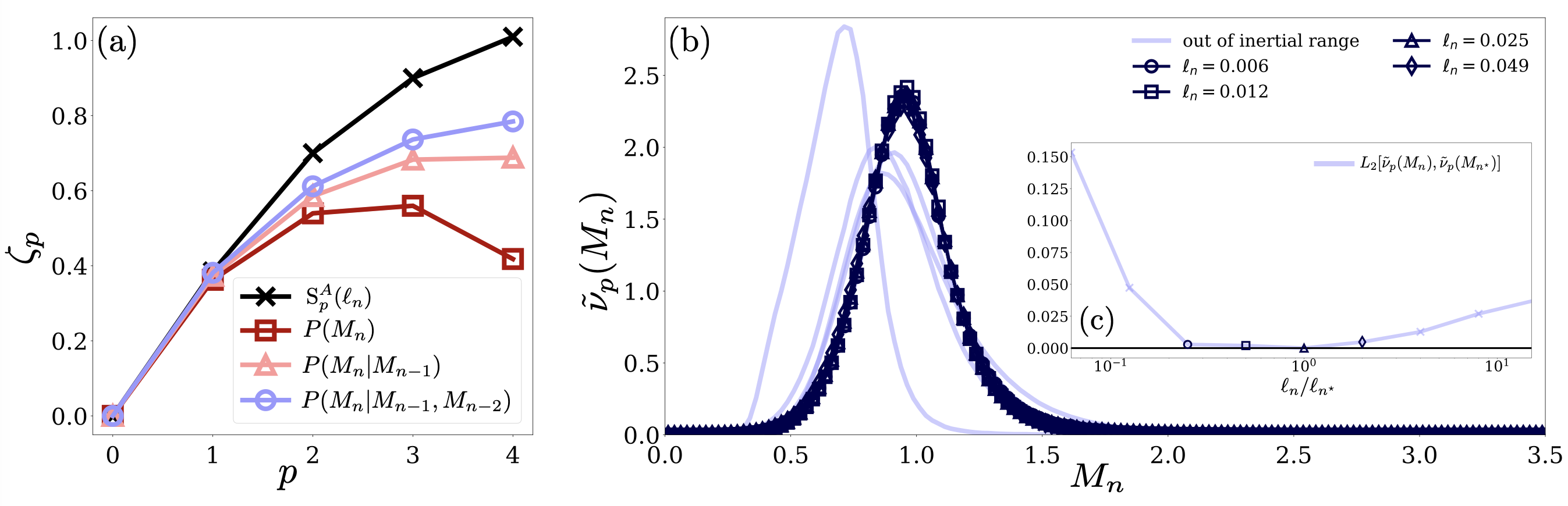}
\caption{(a) Comparison between anomalous exponents $\zeta_p$ for $p=1,\ldots,4$ 
fitted from the structure function ${\rm{S}}^A_p({\ell_n}) =
  \langle A_{\ell_n}^p \rangle$ and predicted by
  the Perron-Frobenious (PF) approach, Eq.\eqref{eq:SF_scaling_PF}, with
  three different approximations of the PF operator: using the measured
  marginal distribution $P(M_n)$ (not including correlations) or the
  conditional distributions $P(M_n|M_{n-1})$ and
  $P(M_n|M_{n-1}, M_{n-2})$, incorporating one- and two-step
  correlations (see legend).  (b) Marginal
  eigenvector measure, $\tilde \nu_p(M_n)$, for $p=4$ and different
  scales.  (c) Distance  in $L_2$ norm between marginal measures at
  different scales with respect to a fixed reference scale  $\ell_{n^\star}$ in the inertial range:  $L_2[\tilde \nu_p(M_n), \tilde
    \nu_p(M_{n^\star})]$.
}
\label{fig:zetap_PerronFrobenius}
\end{figure*}

% \textit{Universality of multipliers statistics.}  
The analogue of
Kolmogorov's multipliers for the original (not rescaled) scalar field
can be defined as \cite{kolmogorov1962refinement,chen2003kolmogorov}:
%\be\label{eq:original_m}
$m_{\ell, \ell'} (\bx, \bold e, \bold e^\prime, t) = |{\delta_{\ell}\theta(\bx, \bold e,  t)}|/|{\delta_{\ell'}\theta(\bx, \bold {e}^\prime, t)}|$,
%\,, \ee
where $\ell$ and $\ell'$ are scales from the inertial interval, and
$\bold e$ and $\bold e^\prime$ are direction vectors. Under the
assumptions of spatial homogeneity and isotropy, the third Kolmogorov
hypothesis \cite{kolmogorov1962refinement} states that the multiplier
statistics are universal and depend only on the scale ratio, $\gamma =
\ell/\ell'$, and the angle between the direction vectors.  By
computing the multipliers in the QL frame, and multiplying and
dividing by the same quantity $A_{\ell}(t)$, we can rewrite the
multipliers as: \be\label{eq:original_m_QL} m_{\ell,
  \ell'}(\bxstar(t),\bold e, \bold e^\prime,t) = \Big|
\frac{\Theta_{\ell}(\bold e, \tau)}{\Theta_{\ell}(\bold
  e^\prime/\gamma, \tau)} \Big|,\,\,\, \ell/\ell' = \gamma\,.  \ee The
hidden scaling invariance of the rescaled field provides theoretical
support to the third Kolmogorov hypothesis, in other words multipliers
statistics should depend only on the scale ratio $\gamma$, which
follows from the statistics of $\Theta_{\ell}(\bold e, \tau)$ being
independent of $\ell$~\footnote{A subtle point in this derivation is
that it is not automatically true that the single-point statistics in
the Eulerian and QL frame coincide. The equivalence of the two
statistics stems from the statistical homogeneity both in the Eulerian
space and the space of initial Lagrangian states $\bx_0 = \bxstar(0)$, together with
incompressibility that ensures that the transformation $\bx_0 \mapsto \bx = \bxstar(t)$ has unit Jacobian at any fixed time.}.

Now, using our numerics, we scrutinize the multiplier statistics across different $\ell$ within the inertial
range, while keeping $\gamma = \ell/\ell'$ fixed. It should be noted
that the PDF of rescaled fields is peaked in $\Theta_{\ell}=0$
(Fig.~\ref{fig:field}b). This heavily influences the tails of PDFs for multipliers defined in \eqref{eq:original_m_QL}, potentially leading to a misleading appearance of universality (see also discussion in
\cite{chen2003kolmogorov}). To mitigate this problem we exploit
Eq.~\eqref{eq:law} and consider multipliers defined in terms of scalar
amplitudes \cite{thalabard2024zero}, which are also expressed in terms
of the rescaled field (Sec.~IV of \cite{Note1})
\be\label{eq:new_mult} M_n := \frac{A_{\ell_n}(t)}{A_{\ell_{n-1}}(t)}
= \langle \Theta^2_{\ell_n}(\lambda\bR,\tau)\rangle^{-1/2}_{\mathcal{D}_1}.
%= \Big(\frac{1}{\pi}\int_{|\bR| \le  1}{\Theta^2_{\ell_n}(\lambda\bR,\tau) d^2\bR} \Big)^{-1/2}.  
\ee
Here, $M_n$ is the multiplier of scalar amplitudes between two
consecutive scales $\ell_n$ and $\ell_{n-1}$, where $\ell_n =
2^{-n}$. The last expression in Eq.(\ref{eq:new_mult}) implies
that universality of these multipliers follows from the HS.
Furthermore, the integral in Eq.(\ref{eq:new_mult}) has a low
probability to be zero. Details on the way we computed the amplitudes $A_{\ell_n}$ can be found in Appendix~B of End Matter.
 Figure~\ref{fig:multpdf}(a) shows the PDF of the
multiplier $M_n$ defined at different scales $\ell_n$, and their
collpse when $\ell_n$ is within the inertial range. Moreover, the
inset of Fig.\ref{fig:multpdf}(b) provides evidence of the
collapse for inertial scales of the one-step conditional distribution
of multipliers, $P(M_n|M_{n-1})$, at least for some values of
$M_{n-1}$.  Indeed from HS, we should expect not only the marginal
distribution $P(M_n)$ to be independent of $n$
(Fig.\ref{fig:multpdf}a), but also that the entire multipliers
statistics is the same provided inertial scales are considered. With
our numerics we could investigate up to the conditional distribution,
$P(M_n|M_{n-1},M_{n-2})$, still verifying the universality (not shown).

%\textit{Perron-Frobenius (PF) scenario to anomalous scaling.} 
% In Fig.\ref{fig:field}(c) we saw that the scalar amplitudes preserve the same scaling exponents of the standard structure functions. Building on this observation, we can associate a measure to the structure functions, ${\rm{S}}_{p}({\ell_n})$. 
In the following, we show how Perron-Frobenius modes of a certain eigenvector measure, which emerges from the statistical hidden scaling symmetry, can be linked to the anomalous exponents.
We start by noticing that the amplitude $A_{\ell_n} = M_n M_{n-1} \cdots M_1 A_{\ell_0}$ is obtained by the telescoping product of multipliers \eqref{eq:new_mult}. We can then
express the respective structure function $\textrm{S}^A_{p}({\ell_n}) \coloneqq \langle A_{\ell_n}^p \rangle = \int d\mu_p^{(n)}$ in terms of the measure
\be\label{eq:measure}
d\mu_p^{(n)} = P^{(n)}(M_n,\dots, M_1) \prod_{k=1,n}M_k^p dM_k \,, \ee
where we fixed the large scale amplitude $A_{\ell_0} = 1$ without lack
of generality.  The joint probability of multipliers
$P^{(n)}(M_n,\dots,M_1)$ connects all scales, from $\ell_n$ in the
inertial range to $\ell_0$ in the forcing range. 
%Since all the terms inside the integral are positive, we can associate a measure, 
%with the structure function $\mathrm{S}_p({\ell_n})$. The measure
%\eqref{eq:measure} extends beyond the inertial range as it involves
%multipliers at the large scales $\ell_0$. 
Consequently, the hypothesis
of statistically restored HS (restricted to the inertial interval only) does not directly apply.  
%to $d\mu_p^{(n)}$, that is why we retained the dependence on $n$ (i.e. on the scale $\ell_n$). 
Using Bayes theorem, we write
$P^{(n)}(M_n,\dots, M_1) = P^{(n)}(M_n | M_{n-1},\dots,
M_1) P^{(n-1)}(M_{n-1},\dots, M_1)$. From this, we see that the
measures (\ref{eq:measure}) at different scales are connected by a
linear operator \be\label{eq:operatorL} \mathcal{L}^{(n)}_p:= M_n^p
P^{(n)}(M_n | M_{n-1},\dots, M_1)dM_n  \ee such that
$d\mu_p^{(n)} = \mathcal{L}^{(n)}_p [d\mu_p^{(n-1)}]$ and, by recursion,
\be\label{eq:iterative_measure} d\mu_p^{(n)} =
\mathcal{L}_p^{(n)} \circ \mathcal{L}_p^{(n-1)} \circ
\mathcal{L}_p^{(n-2)} \circ \dots \circ
\mathcal{L}_p^{(1)}[d\mu_p^{(0)}]\,.  \ee
In principle, every operator depends on both $p$ and $n$. However, this dependence simplifies, i.e. the dependence on $n$ disappears, in the inertial interval, where the HS predicts the universal joint statistics of multipliers. Additionally, Fig.\ref{fig:multpdf}(b) suggests that correlations between multipliers are local, i.e. decay at distant scales, while being noticeable across a few (at least four to five) scales. These arguments yield the inertial-range universality for conditional probabilities, $P^{(n)}(M_n | M_{n-1},\dots, M_1) \to P(M_n | M_{n-1},\dots)$ and, therefore, for the respective operators, $\mathcal{L}_p^{(n)} \to \mathcal{L}_p$. 

The unique operator $\mathcal{L}_p$ governing the
inertial range is a positive linear operator,
mapping positive measure to positive measure. The Perron-Frobenius
theorem ensures a unique maximum eigenvalue $\lambda_p > 0$ for this
operator.  Equation~\eqref{eq:iterative_measure}, reduced to repeated
applications of the same $\mathcal{L}_p$, leads to the asymptotic
form \be\label{eq:asymptotic_dmu} d\mu_p^{(n)} = c_p \lambda_p^n
d\nu_p\,, \ee where $c_p$ is a positive (non universal) coefficient
accounting for large scales statistics and $d\nu_p$ is the eigenvector
measure satisfying $\mathcal{L}_p[d\nu_p] = \lambda_p d\nu_p$ with the normalization $\int d\nu_p =
1$. Integration of (\ref{eq:asymptotic_dmu}) with $\ell_n = 2^{-n}$ yields
\be \label{eq:SF_scaling_PF}
\textrm{S}_p^A(\ell_n) 
%= \!\int\! \!d\mu_p^{(n)} \!
= c_p \lambda_p^n\!
= c_p \ell_n^{\zeta_p}\ \Rightarrow\  \zeta_p \!=\! -\!\log_2
\lambda_p\,. \ee We conclude that
the PF framework expresses the scaling exponents $\zeta_p$ in terms of
the eigenvalues $\lambda_p$.  The PF scenario just described applies
similarly to usual structure functions, ${\rm{S}}_{p}({\ell_n}) = \langle (\delta_{\ell_n}\theta)^p
  \rangle$; see Sec.~V of SM~\cite{Note1}).

%\textit{Numerical validation of the Perron-Frobenius scenario.}  
To numerically verify the PF picture, two natural questions arise: how
to estimate $\lambda_p$, and can we confirm the existence of the
eigenvector measure $d\nu_p$?  To address the first question, we
estimated $\lambda_p$ by considering three successive approximations
of the operator $\mathcal L_p$ in \eqref{eq:operatorL}. Here the full
conditional probability is replaced by $P(M_n)$, $P(M_n|M_{n-1})$ or
$P(M_n|M_{n-1},M_{n-2})$; accounting for further correlations is
computationally prohibitive.  As shown in Sec. VI of SM~\cite{Note1},
the eigenvalues $\lambda_p$ can be computed by solving an iterative
linear problem for marginalized measures $d\mu_p^{(n)}$, with the
results presented in Fig.\ref{fig:zetap_PerronFrobenius}a. Despite the
tested approximations are insufficient, there is a clear trend:
increasing knowledge of correlations improves the estimate. This
suggests that including all essential correlations (up to five or more
scales) may recover the correct scaling.

We now move to the second question, that is to verify the existence of
the eigenvector measure $d\nu_p(M_n,M_{n-1},\dots)$.  Clearly, the
full multi-dimensional verification is impractical. Instead, we check
for the existence of the marginal eigenvector measure, $\tilde
\nu_p(M_n) = \int\nu_p(M_n,M_{n-1},\dots) \, dM_{n-1}\dots$, keeping
only the dependence on $M_n$. From
Eq.(\ref{eq:asymptotic_dmu}) and the normalization $\int d \tilde\nu_p =
1$, we write $\tilde \nu_p(M_n) = \tilde \mu_p^{(n)}(M_n)/\langle A^p_{\ell_n}\rangle$, where the marginal measure $\tilde \mu_p^{(n)}(M_n) =
\langle A^p_{\ell_n} | M_n\rangle P(M_n)$ is expressed from Eq.(\ref{eq:measure}), 
providing direct access to $\tilde \nu_p(M_n)$ from the simulation
data.  The existence of the eigenvector measure $d\nu_p$ guarantees
that its marginal $\tilde \nu_p(M_n)$ is universal within the inertial
range, as confirmed, e.g., for $p=4$ in
Fig.~\ref{fig:zetap_PerronFrobenius}(b) (see also Figs.S1 and S2 in SM~\cite{Note1} for $p=3,5$ and Fig.S3 for the comparison between the marginal eigenvector computed directly via the approximations of the PF operator) showing the collapse at different scales $\ell_n$.
To better quantify the collapse, in Fig.~\ref{fig:zetap_PerronFrobenius}(c) we show the distances between the different curves in
Figs.~\ref{fig:zetap_PerronFrobenius}(b), by fixing a reference
inertial scale, $\ell_{n^\star}$, and computing the $L_2$ norm between
the marginal measure at each scale $\ell_n$ and the one at
$\ell_{n^\star}$. 

%\textit{Conclusions.} 
In this Letter, we have shown that passive
scalars advected by 2D turbulence in the inverse cascade possess a
hidden symmetry. Such HS leads to a number of predictions, namely
universality of the Kolmogorov multipliers (Fig.~\ref{fig:multpdf})
and the link between Perron-Frobenius eigenvalues and anomalous
exponents (Fig.~\ref{fig:zetap_PerronFrobenius}a) and the fact that
its (marginal) eigenvectors are well defined within the inertial range
(Fig.~\ref{fig:zetap_PerronFrobenius}b,c), which have been verified in
high resolution numerical simulations of
Eqs.(\ref{eq:theta}-\ref{eq:NS}).  The hidden symmetry approach
offers an alternative to zero-mode
theory~\cite{falkovich2001particles}.  On the basis of our results
along with previous works on shell
models~\cite{thalabard2024zero,mailybaev2022shell,mailybaev2023hidden} and on NS-equation
  \cite{mailybaev2022hidden,mailybaev2022hidden2} we conjecture that
  HS can offer a unified framework to interpret anomalous
  scaling in hydrodynamic field theory.  This approach overcomes the
  main obstacle, which is the limitation of zero modes to the linear
  setting of turbulent transport. Understanding the links between
  hidden symmetry and zero modes (e.g. in the derivation of fusion rules \cite{eyink1993renormalization,benzi1998multiscale,adzhemyan2001calculation,thalabard2024zero}), with the possibility of extending
  them to NS and other models, revives and opens up new directions for
  research into the theoretical foundations of developed turbulence.

\begin{acknowledgements}
The authors thank Guido Boffetta, Greg Eyink and Massimo Vergassola
for useful discussion. This work was supported by the European
Research Council (ERC) under the European Union’s Horizon 2020
research and innovation programme (Grant Agreement No. 882340). AAM
was also supported by CNPq grant 308721/2021-7, FAPERJ grant
E-26/201.054/2022, and CAPES grant AMSUD3169225P.
\end{acknowledgements}

\begin{appendices}
\section{END MATTER}
\textit{Appendix A: numerical simulations.} This Appendix details the DNS we implemented
\cite{boffetta2000inverse,celani2000universality}. The numerical
integration of the velocity field in Eq.~\eqref{eq:NS} is performed
using the 2D NS equation for vorticity ($\omega=\bm \nabla \times
\bv$): \be\label{eq:w_t} \frac{\partial \omega}{\partial t} +
J(\omega, \Psi) = \nu \Delta^p \omega -\beta \omega - \Delta
f_\omega\,, \ee where $\Psi$ is the stream function, and where the
velocity field is given by $\bv = \bm \nabla^\perp \Psi = (\partial_y
\Psi, -\partial_x \Psi)$. Here, $J(\omega, \Psi)$ represents the
Jacobian determinant. The friction term $-\beta\omega$ extracts energy
from the system at a characteristic friction scale $L_{\beta}\sim
\epsilon^{1/2}\beta^{-3/2}$ with $\epsilon$ the energy flux towards
the large scales, preventing large-scale energy accumulation. We force
at small scales $\ell_{f_\omega}$ with a Gaussian forcing with
correlation function $\langle f_\omega (\bx,t) f_\omega(\bold 0,
t^\prime) \rangle = \delta (t - t^\prime) \mathcal{F}_\omega(|\bx| /
\ell_{f_\omega})$ where $\mathcal{F}(x)$ rapidly decays for $\ell_n
\gg \ell_{f_\omega}$ as $\mathcal{F}_\omega(x) = F_\omega \ell^2_{f_\omega}
\exp(-x^2/2)$. The inertial
range for the inverse cascade of the velocity field is thus
constrained between $\ell_{f_\omega} \ll \ell_n \ll
L_{\textrm{fr}}$. At small scales, enstrophy is dissipated by an
hyperviscous term $\nu (-\Delta)^p \omega$ of order $p=8$, as in
\cite{boffetta2000inverse}.  The advection-diffusion equation
\eqref{eq:theta} is forced at large scale $L_\theta$ with a Gaussian
forcing, ensuring isotropic statistics, with zero mean and a
correlation function $\langle f_\theta(\bx,t) f_\theta(\boldsymbol 0,
t^\prime) \rangle = \delta (t-t^\prime)
\mathcal{F}_\theta(x/L_{\theta})$ and $\mathcal{F}_\theta$ has the
same functional form of $\mathcal{F}_\omega$.  The diffusive term in
Eq.~\eqref{eq:theta} is replaced by a bi-Laplacian operator,
i.e. $\kappa (-\Delta)^2 \theta $.  In particular, we set
$L_{f_\theta} \lesssim L_{\textrm{fr}}$. Additionally, we impose
$\ell_\kappa>\ell_{f_\omega}$, ensuring that the scalar is dissipated
before reaching the forcing scale of the velocity field. These
conditions are chosen to guarantee that the passive scalar dynamics
evolve entirely within the inertial range of the velocity field.  The
numerical integration is carried out using a standard 2/3-dealiased
pseudospectral method on a doubly periodic square domain of size $4096
\times 4096$, with timestepping given by a second-order
Adams–Bashforth scheme. 

\textit{Appendix B: Numerical computation of scalar amplitudes.}
While the integral over the unit disc in Eq.\eqref{eq:A_amplitude} was convenient for the analytical derivation of the hidden symmetry, our theory is equally applicable if the averaging is performed over a different region. In our numerical analysis, we compute the average of scalar fluctuations over $K = 16$ discrete directions. Hence, $A_{\ell}$ is defined as 
\be A_{\ell} = \sqrt{\sum_j [\theta(\mathbf x+ \ell
    \hat{\mathbf{e}}_j) - \theta(\mathbf x)]^2}\,, \ee where the unit
vectors $\hat{\mathbf e}_j$ are chosen as 
\be \hat {\mathbf e}_j =
\Big(\cos \frac{2\pi j}{K},\, \sin \frac{2\pi j}{K} \Big), \quad j = 1,\ldots,K\,.  \ee
Furthermore, for our single-time statistical analysis it is not necessary to switch to the QL frame to collect statistics on $A_{\ell_n}$. Instead, the analysis can be performed directly in the Eulerian frame by averaging over the entire domain. This simplification is justified by the system’s homogeneity and incompressibility, as also true for the original multipliers in Eq.\eqref{eq:original_m_QL}. 
\end{appendices}

%ADD REFERENCES HERE
\bibliographystyle{naturemag}
\bibliography{biblio}

\end{document}

%\linenumbers
\title{Supplementary Material:\\
  Hidden symmetry in passive scalar advected by 2D Navier-Stokes turbulence 
}
%%%%%%%%%%%%%%

\author{Chiara Calascibetta} \email{chiara.calascibetta@inria.fr}
\affiliation{Université Côte d'Azur, Inria, Calisto team, 06902 Sophia Antipolis, France}
%\affiliation{Department of Physics, University of Rome ``Tor
%Vergata'', Via della Ricerca Scientifica 1, 00133 Rome, Italy.} 
%\affiliation{INFN, Sezione di Roma ``Tor Vergata'', Rome, Italy.}
\author{Luca Biferale}
\affiliation{Department of Physics, University of Rome ``Tor
Vergata'', Via della Ricerca Scientifica 1, 00133 Rome, Italy.}
\affiliation{INFN, Sezione di Roma ``Tor Vergata'', Rome, Italy.}
\author{Fabio Bonaccorso}
\affiliation{Department of Physics, University of Rome ``Tor
Vergata'', Via della Ricerca Scientifica 1, 00133 Rome, Italy.}
\affiliation{INFN, Sezione di Roma ``Tor Vergata'', Rome, Italy.}
\author{Massimo Cencini}
\affiliation{Istituto dei Sistemi Complessi, CNR, Via dei Taurini 19, Rome, 00185, Italy.}
\affiliation{INFN, Sezione di Roma ``Tor Vergata'', Rome, Italy.}
\author{Alexei A. Mailybaev}
\affiliation{Instituto de Matemática Pura e Aplicada– IMPA, Rio de Janeiro, Brazil.}

%\date{\today}
\begin{abstract}
\end {abstract}

\maketitle

\begin{figure}[h!]
\centering
\includegraphics[width=0.47\textwidth]{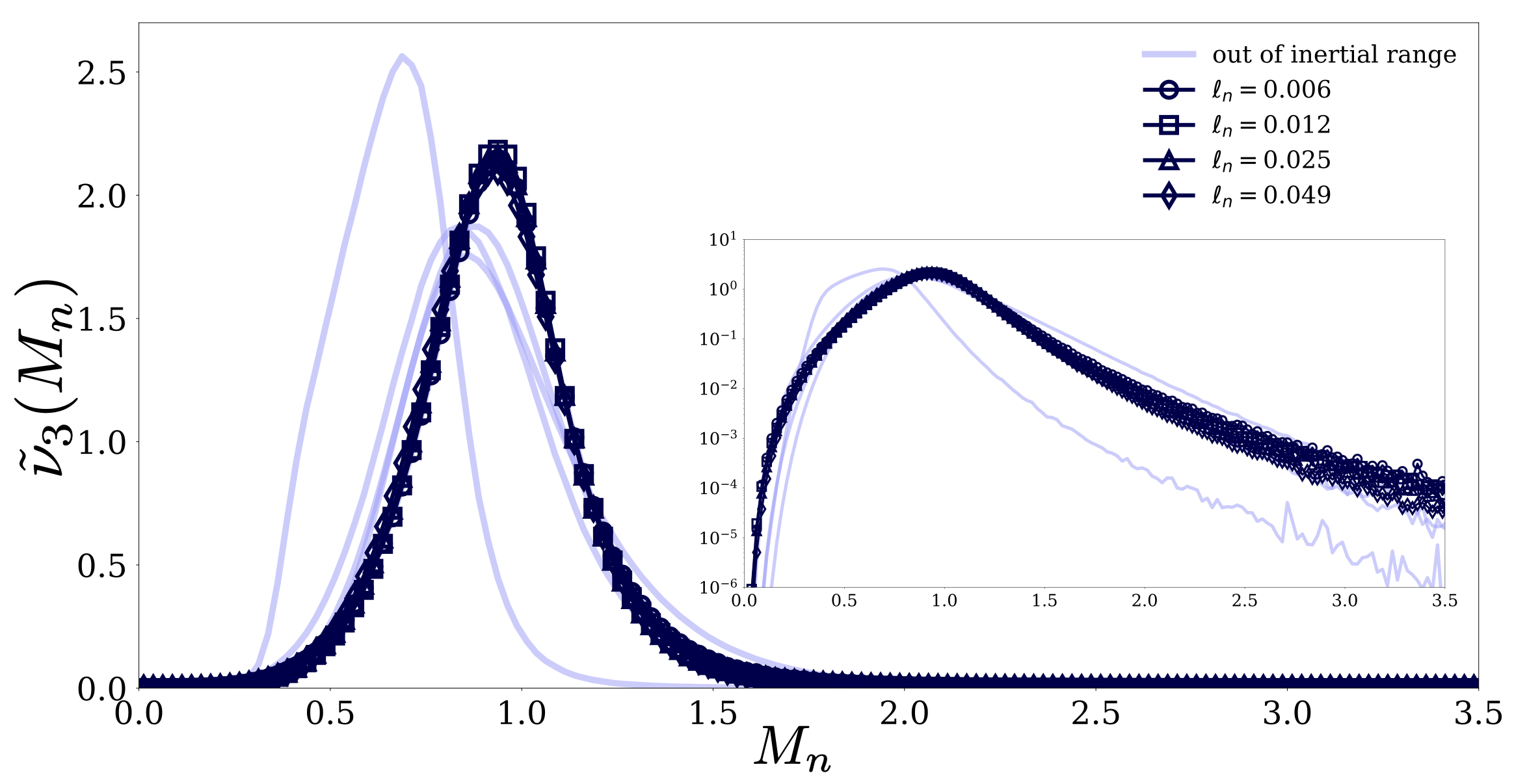}
\caption{Marginal
  eigenvector measure, $\tilde \nu_p(M_n)$, for $p=3$ and different
  scales. In the inset the same plot in y-log scale.}
\label{figSM:PF_p3}
\end{figure}

\begin{figure}[h!]
\centering
\includegraphics[width=0.47\textwidth]{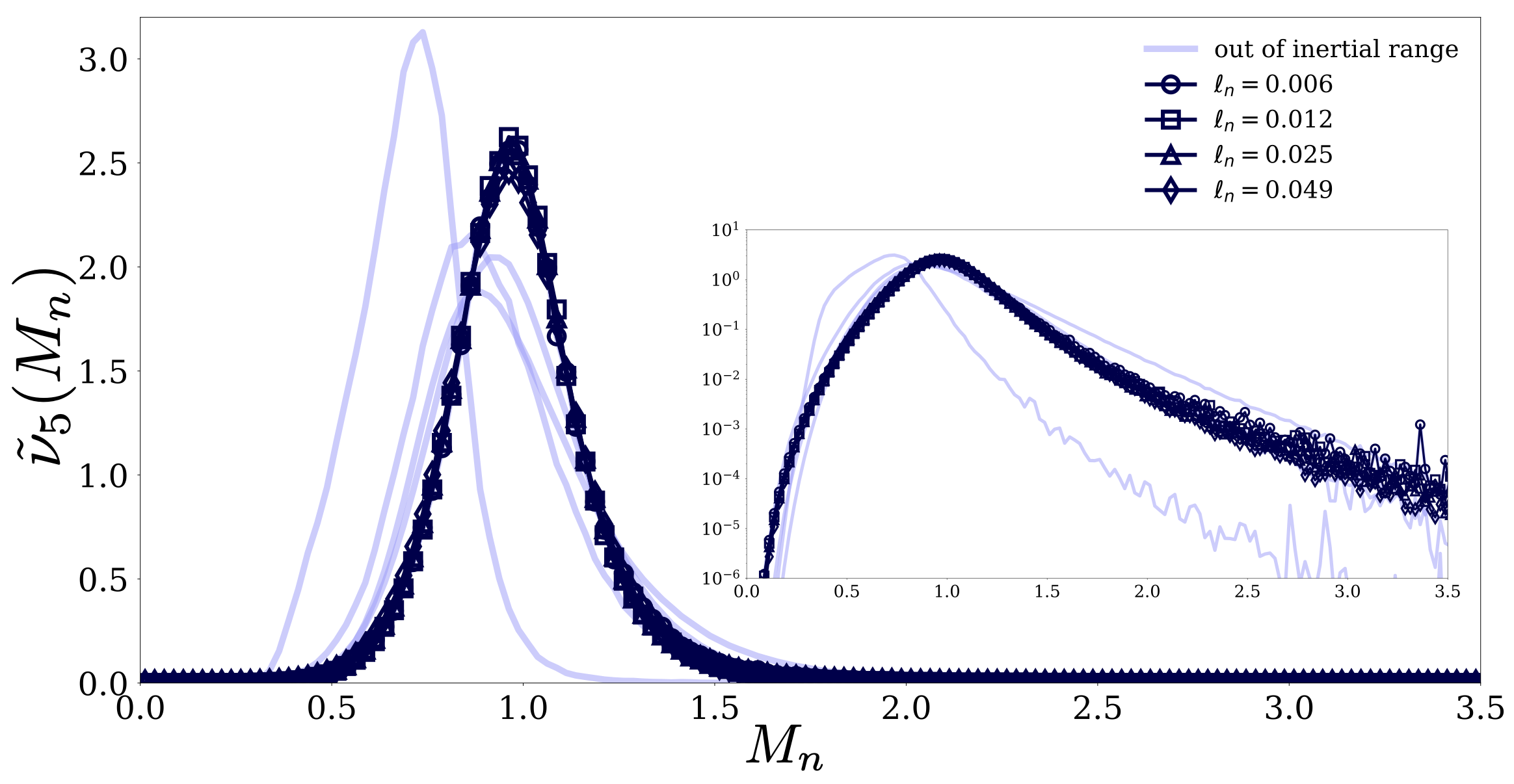}
\caption{Marginal
  eigenvector measure, $\tilde \nu_p(M_n)$, for $p=5$ and different
  scales.  In the inset the same plot in y-log scale.}
\label{figSM:PF_p5}
\end{figure}

\begin{figure}[h!]
\centering
\includegraphics[width=0.47\textwidth]{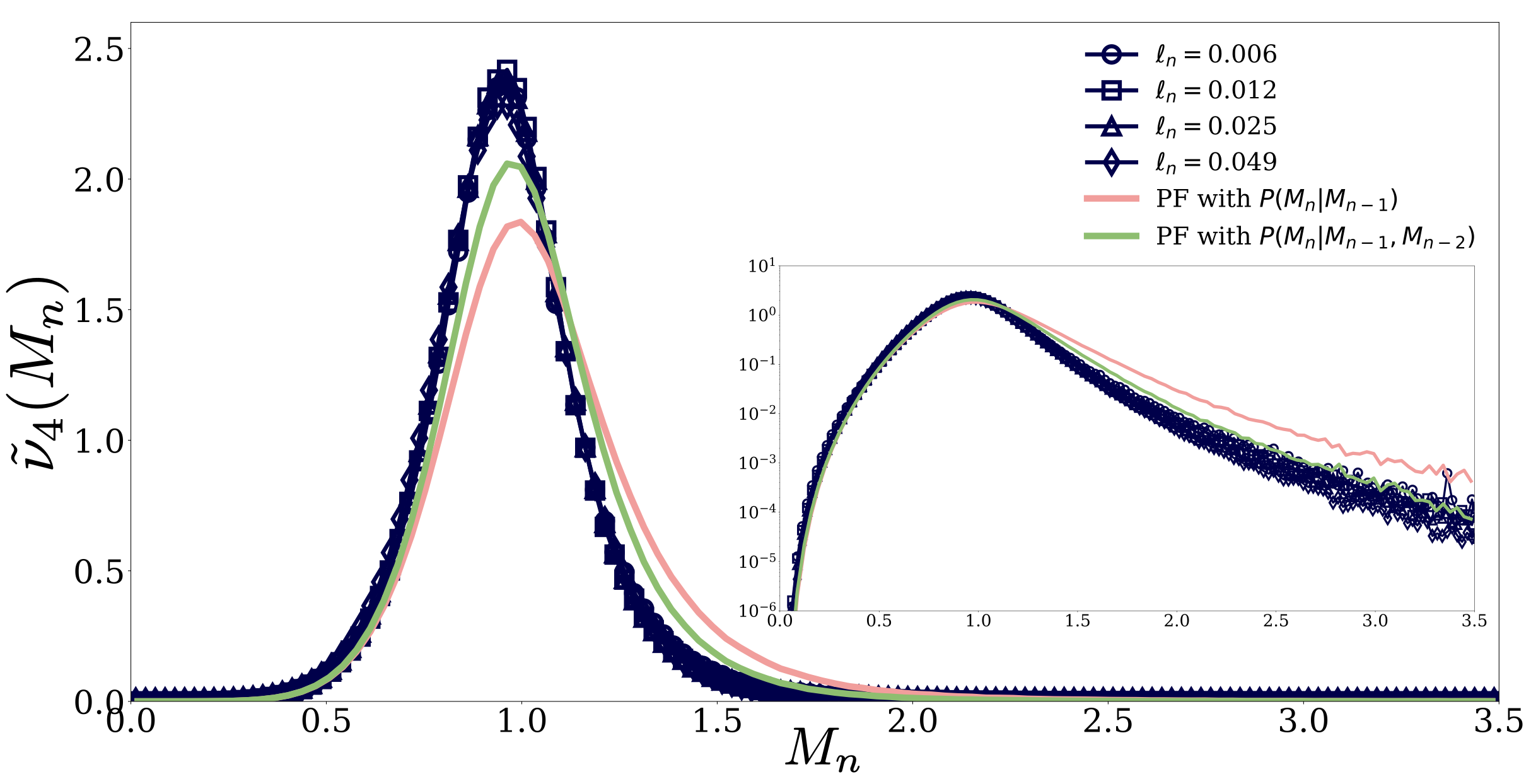}
\caption{Comparison of the eigenvector marginal measures $\tilde \nu_4(M_n)$, derived from the 1-step and 2-step multiplier correlations—$P(M_n|M_{n-1})$ and $P(M_n|M_{n-1}, M_{n-2})$ (represented by red and green curves, respectively, with $\ell_n = 0.025$ fixed)—and from the full statistics (blue curves). In the inset the same plot in y-log scale.}
\label{figSM:PF_approximated}
\end{figure}

This supplementary matter is organized as follows.  First we provide two 
supplementary figures:
Fig.~\ref{figSM:PF_p3} and Fig.~\ref{figSM:PF_p5} that supplement the
results of Fig.3(a) of main text showing the marginal eigenvector
measures, $\tilde \nu_p(M_n)$, for $p=3$ and $p=5$; Fig.~\ref{figSM:PF_approximated} showing how the marginal eigenvector computed by approximating the PF operator approaches $\tilde \nu_p(M_n)$ computed on the simulation data for $p=4$.

Then to ease the
reader we provide a detailed derivation of some of the formulas
presented in the main text, we will use Eq.~(\#) to refer to equations
in the main text and Eq.~(S\#) for equations in the supplementary
material. In particular, in Sec.~\ref{sec:1}  we derive the equation of motion (11) for the rescaled field. In Sec.~\ref{sec:2} we demonstrate that Eq.~(11) is invariant under the change of variable (8) and (12), i.e. we demonstate the hidden  symmetry, which is then shown to correspond to a scaling symmetry in Sec.~\ref{sec:3}.
Sec.~\ref{sec:4} provides a derivation of Eq.~(15), i.e. demostrates the link between multipliers and rescaled fields.
Finally, in  Secs.~\ref{sec:5} and \ref{sec:6} we show how to apply the PF approach for the standard structure functions and discuss the PF scenario detailing the approximations we used for the PF operator and the way the eigenvalue can be iteratively found.

\section{Derivation of Eq.~(11): rescaled equations for the ideal passive scalar system \label{sec:1}}
In this section we present a step-by-step derivation of Eq.~(11).
Starting from Eq.~(10)  with $t = t_{\ell} \tau$, the chain rule yields
\be\label{eq:deriv_Thetaeq0}
\begin{gathered}
    \partial_\tau \Theta_{\ell} 
    = 
    \frac{t_{\ell}}{A_{\ell}(t)}
    \frac{\partial}{\partial t}\delta_{\ell} \theta 
    -t_{\ell} \,
      \frac{\delta_{\ell} \theta}{A^2_{\ell}(t)}  
      \frac{d A_{\ell}}{dt} ,
\end{gathered}
\ee
where we used the shorthand $\delta_{\ell} \theta = \theta(\bxstar(t) + \ell \bR,t) - \theta(\bxstar(t),t)$, notice that $A_\ell$ depends only on time.
Then we consider the first term of the above equations, which can be written as
\be\label{eq:deriv_Thetaeq1}
\begin{gathered}
    \frac{\partial}{\partial t}\delta_{\ell} \theta 
    = 
    \partial_t\theta(\bxstar(t) + \ell \bR,t)
    -\partial_t\theta(\bxstar(t),t)
    \\[2pt]
    +\frac{d\bxstar}{dt} \cdot \nabla\theta(\bxstar(t) + \ell \bR,t)
    -\frac{d\bxstar}{dt} \cdot
    \nabla\theta(\bxstar(t),t).
\end{gathered}
\ee
Here the expressions $\partial_t\theta(\bxstar(t) + \ell \bR,t)$ and $\nabla\theta(\bxstar(t) + \ell \bR,t)$ denote the derivatives $\partial_t\theta$ and $\nabla\theta$ evaluated at $\bx = \bxstar(t) + \ell \bR$.
Similarly we express
\be \label{eq_AppC_1D}
\nabla_{\bR} \Theta_{\ell} 
= \frac{\ell}{ A_{\ell}(t) } \nabla\theta(\bxstar(t) + \ell \bR,t).
\ee 

Using the ideal transport equation $\partial_t\theta = -\bv \cdot \nabla \theta$ and the definition $d\bxstar/dt = \bv(\bxstar(t),t)$ of the Lagrangian tracer, Eq.(\ref{eq:deriv_Thetaeq1}) after a cancellation of two terms becomes
\be\label{eq:deriv_Thetaeq1step2}
\begin{gathered}
    \frac{\partial}{\partial t}\delta_{\ell} \theta 
    = -\bv(\bxstar(t) + \ell \bR,t) 
    \cdot \nabla\theta(\bxstar(t) + \ell \bR,t)
    \\
    +\, \bv(\bxstar,t)
    \cdot \nabla\theta(\bxstar(t) + \ell \bR,t).
\end{gathered}
\ee
From definitions (5) and (6) we have
\be\label{eq:deriv_Thetaeq1C}
\bv(\bxstar(t) + \ell \bR,t) - \bv(\bxstar(t),t) 
= \frac{\ell}{t_{\ell}} \bV_{\ell}(\bR,\tau).
\ee
Using Eqs.(\ref{eq_AppC_1D}) and (\ref{eq:deriv_Thetaeq1C}), 
we reduce Eq.(\ref{eq:deriv_Thetaeq1step2}) to the form
\be\label{eq:deriv_Thetaeq1step3}
\begin{gathered}
    \frac{t_{\ell}}{A_{\ell}(t)}
    \frac{\partial}{\partial t}\delta_{\ell} \theta 
    = -\bV_{\ell}
    \cdot \nabla_{\bR} \Theta_{\ell}.
\end{gathered}
\ee

The time derivative of $A_{\ell}(t)$ is expressed from the definition (9) as
\be\label{eq:A_timedev}
\begin{gathered}
\frac{d A_{\ell}}{d t} = \frac{1}{A_{\ell}(t)} \Big\langle \delta_{\ell} \theta \, \frac{\partial}{\partial t}\delta_{\ell} \theta\Big\rangle_{\mathcal D_1}.
\end{gathered}
\ee
Using Eq.(\ref{eq:deriv_Thetaeq1step3}) and expressing $\delta_{\ell} \theta/A_{\ell}(t) = \Theta_{\ell}(\bR,\tau)$ from Eq.~(10), we find
\be\label{eq:A_timedevB}
\begin{gathered}
\frac{t_{\ell}}{A_{\ell}(t)}
\frac{d A_{\ell}}{d t} 
= -\big\langle \Theta_{\ell} \bV_{\ell}
    \cdot \nabla_{\bR} \Theta_{\ell} 
    \big\rangle_{\mathcal D_1}.
\end{gathered}
\ee
Using expressions (\ref{eq:deriv_Thetaeq1step3}), (\ref{eq:A_timedevB}) and Eq.~(10)  in Eq.(\ref{eq:deriv_Thetaeq0}) yields Eq.~(11).

As a final remark, one should notice that while the original equation for the passive scalar (and equivalently its QL-version for the scalar increments) was linear, the equation for the rescaled field (i.e. Eq.~(11)) is non-linear. Still it enjoys of the hidden scaling symmetry, as derived in the next secttion. 

\section{Demonstration that Eq.(11) is invariant under the change of variable (8) and (12): derivation of the hidden symmetry}\label{sec:2}
In this section we prove that the transformation in Eqs.(12) and (8) is indeed a symmetry for the rescaled passive scalar system, Eq.~(11). Assuming that $\bV(\bR,\tau)$
and $\Theta(\bR,\tau)$ satisfy Eq.~(11), we have to
show that the transformed fields \be
	\label{eq_AppD_der1}
	\begin{gathered}
	\bV'(\bR,\tau) = \gamma^{1/3}\bV(\bR/\gamma,\tau/\gamma^{2/3}),
	\\[2pt]
	\Theta'(\bR,\tau) 
	= \frac{\Theta(\bR/\gamma,\tau/\gamma^{2/3})}{B_\gamma(\tau)}
	\end{gathered}
	\ee
solve the same equation, i.e.,
	\be
	\label{eq_AppD_der1b}
	\partial_\tau \Theta' + \bV'\cdot \nabla_{\bR}\Theta' 
	= \Theta' \langle \Theta' \bV' \cdot \nabla_{\bR}\Theta'\rangle_{\mathcal{D}_1}\,.
	\ee 
Substituting Eq.(\ref{eq_AppD_der1}) into the lhs of Eq.(\ref{eq_AppD_der1b}), after
differentiation and dropping the common factor we have
	\be
	\label{eq_AppD_der2}
	\frac{\Theta_\tau+\bV\cdot \nabla_{\bR}\Theta}{\gamma^{2/3}B_\gamma(\tau)}
	-\frac{\Theta}{B^2_\gamma(\tau)}\frac{dB_\gamma}{d\tau} 
	= \frac{\Theta \langle \Theta \bV \cdot \nabla_{\bR}\Theta\rangle_{\mathcal{D}_1}}{\gamma^{2/3}B^3_\gamma(\tau)},
	\ee
where all fields in the numerators are evaluated at $(\bR/\gamma,\tau/\gamma^{2/3})$. 
Using Eq.(12) in the first term and then multiplying by $-\gamma^{2/3}B^3_\gamma/\Theta$, we further write
	\be
	\label{eq_AppD_der2c}
	\gamma^{2/3}B_\gamma(\tau) \frac{dB_\gamma}{d\tau} 
	+\left(1-B^2_\gamma(\tau)\right)
	\langle \Theta \bV \cdot \nabla_{\bR}\Theta\rangle_{\mathcal{D}_1} = 0.
	\ee
Expressing $B_\gamma(\tau) =\langle \Theta^2(\bR/\gamma,\tau/\gamma^{2/3})\rangle^{1/2}_{\mathcal{D}_1}$, we have
	\be
	\label{eq_AppD_der2d}
	\langle \Theta \partial_\tau \Theta\rangle_{\mathcal{D}_1}+
	\left(1-\langle \Theta^2 \rangle_{\mathcal{D}_1}\right)
	\langle \Theta \bV \cdot \nabla_{\bR}\Theta\rangle_{\mathcal{D}_1} = 0,
	\ee
recalling that all fields assume the argument $(\bR/\gamma,\tau/\gamma^{2/3})$. 
This equation coincides with Eq.(11) multiplied by $\Theta$ and averaged over the disc $\mathcal{D}_1$, therefore, proving Eq.(\ref{eq_AppD_der1b}).

\section{Demonstration that the hidden symmetry corresponds to a scaling symmetry}\label{sec:3}

Now let us show that the hidden symmetry relates the two scales, $\ell$ and $\ell' = \ell / \gamma$. Namely, we will prove that the relations (12) transform the respective rescaled field $\Theta_{\ell}(\bR,\tau)$ to $\Theta_{\ell'}(\bR^\prime,\tau^\prime)$, where the primes remind that the rescaled variables (4) depend on the reference scale $\ell'$. 
Expressing 
\be
\label{eq:App_dlt}
\delta_{\ell'}\theta(\bxstar(t),\bR^\prime,t) = \delta_{\ell}\theta(\bxstar(t),\bR^\prime/\gamma,t)
\ee
from Eq.(3), we write the amplitude (9) at scale $\ell'$ as
\be\label{eq:A_amplitudePr}
\begin{gathered}
A_{\ell'}(t) = \Big(\frac{1}{\pi}\int_{|\bR'| \le 1} \big[\delta_{\ell} \theta(\bxstar(t),\bR'/\gamma,t)\big]^2 d^2\bR' \Big)^{1/2}\,.
\end{gathered}
\ee
Dividing by $A_{\ell}(t)$ and recalling definition (10) yields
\be\label{eq:A_amplitudeRatio}
\begin{gathered}
\frac{A_{\ell'}(t)}{A_{\ell}(t)} = \Big(\frac{1}{\pi}\int_{|\bR'| \le 1}\Theta_{\ell}^2(\bR'/\gamma,\tau) d^2\bR' \Big)^{1/2}\,,
\end{gathered}
\ee
where $\tau = t/t_{\ell}$.
Similarly, using the definition (10) and the identity (\ref{eq:App_dlt}), we express
\be\label{eq:Theta_HS_deriv1}
\begin{gathered}
\Theta_{\ell'}(\bR^\prime,\tau^\prime) 
% = \frac{\delta_{\ell'}\theta(\bxstar(t),\bR^\prime,t)}{A_{\ell_{n}}(t)}
%\\[2pt]
= \frac{\delta_{\ell}\theta(\bxstar(t),\bR^\prime/\gamma,t)}{A_{\ell}(t)} 
\frac{A_{\ell}(t)}{A_{\ell'}(t)},
%= \Theta_{\ell_{n}}(\bR^\prime/\gamma,\tau) \, \frac{A_{\ell_{n}}(t)}{A_{\ell_{n^\prime}}(t)},
\end{gathered}
\ee
where also multiplied and divided by the same quantity $A_{\ell}(t)$. Using  Eqs.(10) and  (\ref{eq:A_amplitudeRatio}) in Eq.\eqref{eq:Theta_HS_deriv1} we derive
\be\label{eq:Theta_HS_deriv_fin}
\begin{gathered}
    \Theta_{\ell'}(\bR^\prime,\tau^\prime) 
    = \frac{\Theta_\ell(\bR'/\gamma,\tau'/\gamma^{2/3})}{
    \Big(\frac{1}{\pi}\int_{|\bR'| \le 1}{\Theta_\ell^2(\bR'/\gamma,\tau'/\gamma^{2/3}) d^2\bR'} \Big)^{1/2}
    }\, ,
\end{gathered}
\ee
where we also substituted $\tau = \tau'/\gamma^{2/3}$ following from the scaling relations (4) and (5). This is the hidden symmetry transformation (12). 

\section{Derivation of Eq.~(15)}
\label{sec:4}

Relation (15) follows from Eq.(9) and  (10) as
\be\label{eq:new_mult_App}
\begin{gathered}
M_n 
= \left(\frac{A_{\ell_{n-1}}(t)}{A_{\ell_{n}}(t)}\right)^{-1}
\\[2pt]
= \Big(\frac{1}{\pi A^2_{\ell_{n}}(t)}\int_{|\bR| \le 1}{\big[\delta_{\ell_{n-1}}\theta(\bxstar(t),\bR,t)\big]^2 d^2\bR} \Big)^{-1/2}
\\[4pt]
= \Big(\frac{1}{\pi}\int_{|\bR| \le 1}{\bigg[\frac{\delta_{\ell_{n}}\theta(\bxstar(t),\lambda\bR,t)}{A_{\ell_{n}}(t)}
\bigg]^2 d^2\bR} \Big)^{-1/2}
\\[4pt]
= \Big(\frac{1}{\pi}\int_{|\bR| \le 1}{\Theta^2_{\ell_n} (\lambda\bR,\tau) d^2\bR} \Big)^{-1/2},
\end{gathered}
\ee
where we also used Eq.(3) with $\ell_{n-1} = \lambda\ell_n$.

%}endForestGreen

%{\color{ForestGreen}
\section{Perron-Frobenius scenario for other structure functions}\label{sec:5}
Here we show that the Perron--Frobenius approach can be extended to other definitions of the structure functions, including the usual one in terms of scalar increments from Eq.~(3), namely, the moments $\langle |\delta_\ell \theta(\bx,\mathbf{e},t)|^p \rangle$. We write these increments in terms of the multipliers (15) and the rescaled field (10) as
\be
|\delta_{\ell_n} \theta(\bxstar(t),\mathbf{e},t)|^p = |\Theta_{\ell_n}(\mathbf{e},\tau)|^p M_n^p \cdots M_1^p A_{\ell_0}^p.
\ee
Using the measure (16), we express the respective structure function as
\be
\big\langle |\delta_{\ell_n} \theta|^p \big\rangle 
= \int Q^p P_Q^{(n)}(Q|M_n,M_{n-1},\ldots)\, 
d Q\, d\mu_p^{(n)},
\ee
where $P_Q^{(n)}(Q|M_n,M_{n-1},\ldots)$ is the conditional probability density of $Q = |\Theta_{\ell_n}(\mathbf{e},\tau)|$.  Assuming that $P_Q^{(n)}$ is scale-local (thus, depending only on the scales from the inertial interval), the hidden scaling symmetry ensures that $P_Q^{(n)} = P_Q$ does not depend on $n$. Finally, using the Perron--Frobenius asymptotics (19), similarly to Eq.(20) we derive
\be
\begin{gathered}
\big\langle |\delta_{\ell_n} \theta|^p \big\rangle 
= q_p c_p \ell_n^{\zeta_p}\,, \quad \zeta_p = -\log_\lambda \lambda_p,
\end{gathered}
\ee
where $q_p = \int Q^P P_Q(Q|M_n,M_{n-1},\ldots)\, dQ \, d\nu_p$. We obtained the scaling law with the same exponent $\zeta_p$ but a different constant prefactor $q_p$.
%}endForestGreen

\section{Perron-Frobenius Scenario}\label{sec:6}
In this Appendix, we discuss different approximations for estimating
the Perron-Frobenius eigenvalue $\lambda_p$ and investigate the
relationship $\zeta_p = -\log_\lambda \lambda_p$ presented in Fig.3(a)
of the main text. The main focus is on how to derive marginal measures
using a stepwise approach in the context of multipliers statistics.
As for the notation, we rewrite the measure in Eq.~(16)
as $d\mu_p^{(n)} = \mu_p^{(n)}dM_n\dots dM_1$.

%{\color{ForestGreen}
We start with the uncorrelated case, where we assume that the variables $M_n$ for different $n$ are independent with the same probability distribution $P(M_n)$. In that case, $\textrm{S}^A_{p}({\ell_n}) \coloneqq  \int d\mu_p^{(n)}$ yields 
\begin{equation}\label{eq:SF_Aln_ind}
\begin{gathered}
    \textrm{S}^A_{p}({\ell_n}) = c_p\langle M^p\rangle^n, \quad \langle M^p\rangle = \int M^p P(M) dM,
\end{gathered}
\end{equation}
where $\langle M^p\rangle$ is a moment of every multiplier in the inertial interval and the coefficient $c_p$ accounts for deviations from the universality in the forcing range. We recover the scaling (20) with $\lambda_p = \langle M^p\rangle$.

Next we consider the 1-step conditional distribution $P(M_n|M_{n-1})$, which is an approximation of the full conditional distribution. This allows us to define the marginal measure $\tilde \mu_p^{(n)}(M_n)$ as follows:
\be\label{eq:appendix_PF1}
\begin{gathered}
    \tilde \mu_p^{(n)}(M_n) 
 = \int \mu_p^{(n)}dM_{n-1}\dots dM_1 = \\
= \int M_n P(M_n|M_{n-1}) d\mu_p^{(n-1)}\,,
\end{gathered}
\ee
where we have skipped the integration by $dM_n$ and where we have assumed $P(M_n|M_{n-1},\cdots) \equiv P(M_n|M_{n-1})$. The marginal measure at the $n$-th scale, $\tilde \mu_p^{(n)}$, is related to the marginal measure  $\tilde \mu_p^{(n-1)}$ by:
\be\label{eq:recursive_tildemu}
\begin{gathered}
 \tilde \mu_p^{(n)}(M_n)
= \int M_n^p P(M_n | M_{n-1}) \tilde \mu_p^{(n-1)}(M_{n-1}) dM_{n-1}\,.
\end{gathered}
\ee
As $n\to \infty$, we expect the marginal measure to collapse to an eigenvector-like form:
\be
\tilde \mu_p^{(n)}(M_n) = c_p \lambda_p^n \tilde \nu_p(M_n)\,,
\ee
where $c_p$ is a constant and $\tilde \nu_p(M_n)$ is the marginal eigenvector measure associated with the eigenvalue $\lambda_p$. 
Substituting this into the recursive relation in Eq.\eqref{eq:recursive_tildemu}, we get the following linear problem:
\be
\lambda_p \tilde \nu_p(M_n) = \int M_n^p P(M_n|M_{n-1}) \tilde \nu_p(M_{n-1}) dM_{n-1}\,.
\ee
This is a standard eigenvalue problem, which can be solved numerically for $\lambda_p$.

Finally, we consider the case of a 2-step conditional distribution, where $P(M_n|M_{n-1},M_{n-2})$ is used instead of $P(M_n|M_{n-1})$. The corresponding marginal measure statisfies the following relation:
\be\label{eq:2-step_PF}
\begin{gathered}
\lambda_p \tilde \nu_p(M_n, M_{n-1}) = \\ =  \int M_n^p P(M_n|M_{n-1},M_{n-2}) \tilde \nu_p(M_{n-1}, M_{n-2}) dM_{n-2}\,.
\end{gathered}
\ee
This equation is an iterative linear problem to be solved numerically for $\lambda_p$, where the marginal measure $\tilde \nu_p(M_n, M_{n-1})$ depends on two variables. 
%}endForestGreen

%\section{Appendix to be added my alexei related to inversion of formula for the marginal measure}

% Now, let us consider the uncorrelated case, where we assume that the variables $M_n$ and $M_{n-1}$ are independent. In this case, we have $P(M_n|M_{n-1},M_{n-2}) = P(M_n)$, and the marginal measure $\tilde \nu_p(M_n, M_{n-1})$ can be rewritten as the product of the individual marginal distributions. This leads to the following expression:
% \be
% \tilde \nu_p(M_n, M_{n-1}) = M_n^p M_{n-1}^p P(M_n) P(M_{n-1})\,.
% \ee
% Substituting this into the 2-step equation \eqref{eq:2-step_PF}, we obtain:
% \be
% \begin{gathered}
% \lambda_p M_n^p P(M_n) M_{n-1}^p P(M_{n-1}) = \\ =  \int M_n^p P(M_n) M_{n-1}^p M_{n-2}^p  P(M_{n-1}) P(M_{n-2}) dM_{n-2}\,.
% \end{gathered}
% \ee
% The integral simplifies to $\int M_{n-2}^p P(M_{n-2})dM_{n-2} = \langle M_{n-2}^p \rangle$. Thus, the eigenvalue $\lambda_p$ in the uncorrelated case can be expressed as 
% $\lambda_p = \langle M_n^p \rangle$\,, where $n$ is a generic index substituting $n-2$ because of universality of multipliers statistics in the inertial interval.  

%ADD REFERENCES HERE
\bibliographystyle{naturemag}
\bibliography{biblio}